\documentclass[twocolumn, amsmath, amssymb,10pt]{extarticle}

\usepackage{graphicx}
\usepackage{subfigure}
\usepackage{xcolor}
\usepackage{physics}
\usepackage{bm}
\usepackage{enumerate}
\usepackage{siunitx}
\usepackage{float}
\usepackage{lipsum}
\usepackage{epsfig}
\usepackage{svg}
\usepackage{dsfont}
\usepackage{amsfonts}
\usepackage[numbers, square]{natbib}
\usepackage{geometry}
\usepackage{tikz}
\usepackage[resetlabels,labeled]{multibib}
\newcites{S}{\vspace{0pt}}

\usepackage[font=small,figurename=FIG.]{caption}
\usepackage{authblk} 
\usepackage{abstract}

\usepackage[colorlinks, linkcolor=blue, citecolor=blue, urlcolor=blue]{hyperref}
\usepackage{cleveref}

\begin{document}
\pagenumbering{gobble}
\twocolumn[
\begin{@twocolumnfalse}
\begin{center}
  \textbf{\large Phase-Biased Andreev Diffraction Grating}\\ \vspace{0.4cm}
  \small Magnus~R.~Lykkegaard,$^{1}$ Anders~Enevold~Dahl,$^{1}$ Karsten~Flensberg,$^{1}$ \\ \small Tyler~Lindemann,$^{2,3}$ Michael~J.~Manfra,$^{2,3,4,5}$ and Charles~M.~Marcus$^{1,6,7}$\\ \vspace{0.1cm}
  {\small\itshape ${}^1$Center for Quantum Devices, Niels Bohr Institute, \\ University of Copenhagen, DK-2100 Copenhagen, Denmark\\
  ${}^2$Department of Physics and Astronomy, Purdue University, West Lafayette, Indiana 47907, USA\\
  ${}^3$Birck Nanotechnology Center, Purdue University, West Lafayette, Indiana 47907, USA\\
  ${}^4$School of Electrical and Computer Engineering,\\ Purdue University, West Lafayette, Indiana 47907, USA\\
  ${}^5$School of Materials Engineering, Purdue University, West Lafayette, Indiana 47907, USA\\
  ${}^6$Department of Physics, University of Washington, Seattle, Washington 98195, USA\\
  ${}^7$Materials Science and Engineering, University of Washington, Seattle, Washington 98195, USA\\}
\small(Dated: \today)\\ \vspace{0cm}
\end{center}

\begin{center}
\hspace*{-1cm}
\begin{minipage}{\dimexpr\textwidth-2.5cm}
\begin{abstract}
In optical diffraction, the phase difference between sources in a grating or multi-slit mask is determined by the angle to the imaging screen, yielding the familiar  multi-lobed diffraction image. Here, we realize a similar phenomenon in a superconductor-semiconductor hybrid circuit configured to allow Andreev scattering from multiple parallel scatterers. Phase differences between scatterers are set by tapping off of a remote superconducting meander. We investigate arrays with two, three, four, and ten Andreev scatterers, examining local and nonlocal diffraction patterns, finding good agreement with a theory of multiple Andreev scattering, not to be confused with multiple Andreev reflection.  Adding current-carrying taps to the meander allows individual phase control.
\end{abstract}
\end{minipage}
\end{center}

\end{@twocolumnfalse}
]
An electron impinging on a superconductor (S) from a normal metal (N) may undergo Andreev reflection, which phase-conjugates the electron wave function  resulting in a retroreflected hole, similar to how light is reflected by a phase-conjugate mirror \cite{PhaseConjMirror}. 
With multiple superconducting interfaces, retroreflected electron and holes can interfere either destructively or constructively depending on the phase difference between superconducting elements. The superconducting phase is readily observed, for instance, in circuits containing Josephson junctions (JJs) configured as superconducting quantum interference devices (SQUIDs). SQUIDs are widely used to measure small magnetic fields \cite{SQUID1,SQUID2,SQUID3}, as well as to modulate coupling in superconducting qubits \cite{Krantz.2019, Blais.2021, Rasmussen.2021}, and, more recently, to investigate fundamental processes in Andreev molecules \cite{AMS1,AMS2}. 

\begin{figure}[h] 
\centering
\includegraphics[width=0.43\textwidth]{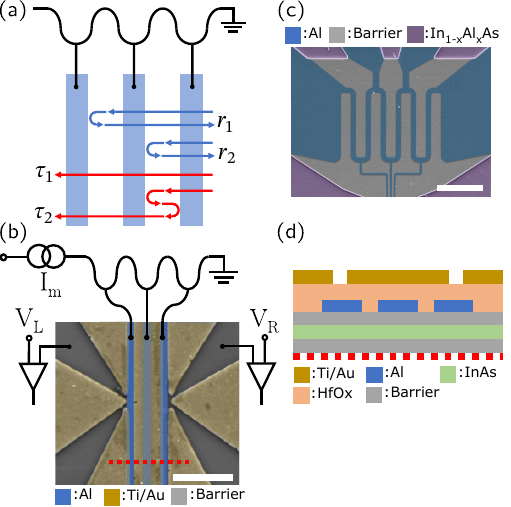}
\caption{(a) Schematic of the three-wire ($M=3$) device, with superconducting wires (light blue) connected to a remote superconducting meander. Multiple phase-sensitive transmissions, $\tau_i$, and retroreflections, $r_i$, contribute to  overall local and nonlocal conductances. (b) False-color micrograph showing three-wire device and circuitry.  Scale bar is $1$ $\mu$m. (c) Zoomed-out false-color micrograph showing the Al meander with three downward phase taps and two upward current taps. Regions beyond the mesa are deep-etched to the In$_{\rm 1-x}$Al$_{\rm x}$As graded buffer layer (See SM \cite{sup}). Scale bar is $10$ $\mu$m. (d) Cross section along red dashed line in (b), showing heterostructure and fabrication layers. The InAs quantum well is 7 nm with In$_{0.75}$Ga$_{0.25}$As barriers (10 nm top, 4 nm bottom) (See SM Sec~\ref{sup:S0} \cite{sup}). \label{fig1}} 
\end{figure}

In this Letter, we extend the phase-biasing scheme used in  Andreev molecules to realize structures with multiple retroreflecting  superconductor-semiconductor interfaces, creating an {\it Andreev diffraction grating} (ADG) using a patterned superconductor-semiconductor heterostructure. Denoting these structures as diffraction gratings emphasizes the analogy to optics~\cite{Cheraghci.2016} and electronic optical analogs~\cite{Graphene.2024}, now including multiple Andreev processes, and using a  tapped superconducting line to set phases.
The analogy between multi-slit interference and multiple junctions in parallel has been noted previously~\cite{Courtois.1995,DeLuca.2015}.  
Also, because the ADG is retroreflecting, there is a superficial similarity to so-called metagratings in optics~\cite{Metagrating1,Metagrating2}. 
In contrast to metagratings, however, each scattering event in the ADG is retroreflective, rather than retroreflection emerging from a collection of normal scatterers. Previous related work used superconductor-insulator-superconductor (SIS) JJs in multiple loops, where interference was controlled with an applied magnetic field~\cite{M.Lucci.2018}. We instead realize phase-tuning of the grating using regularly spaced taps off of a remote superconducting meander line, without external magnetic fields.  We have investigated devices with two, three, four, and ten parallel superconducting reflectors, generalizing earlier work where two reflectors were controlled by an applied current \cite{CurrentControl_1995}.  A device with three reflectors is shown in Fig.~\ref{fig1}. We develop a theoretical model of the ADG based on multiple Andreev reflection and interference, and find good qualitative agreement between experiment and theory. 
Devices were fabricated on an InAs heterostructure on an InP wafer, grown by molecular beam epitaxy, with a 5 nm epitaxial Al layer on the surface of a 7 nm InAs quantum well with a 10 nm In$_{0.75}$Ga$_{0.25}$As top barrier. 
Normal (non-superconducting) regions were formed by selectively removing the Al top layer with Transene D wet etch, with each device on a separate mesa, defined by deep etching into the graded buffer layer below the quantum well and barriers. A micrograph of a three-wire device is shown in Fig.~\ref{fig1}(b) along with a cross-section of the top of the material stack [Fig.~\ref{fig1}(d)]. The Au/Ti gates (gold color) were either operated as constriction gates that prevented conduction beyond the ADG or to control the carrier density in the normal regions between superconducting wires. We voltage bias the gray area on either side, and measure local and non-local conduction using lock-in amplifiers (see Fig.~\ref{figsup:6} for detailed circuitry.)
Gates were isolated from the superconducting and normal regions with $\mathrm{HfO_{2}}$ deposited by atomic layer deposition. Located on either side of the ADG, the constriction gates are separated into three gates for increased controllability. Voltage on the plunger gate, the gate on top of the grating, was kept less negative than $-0.3$~V. Voltage on constriction gates were kept less negative than $-0.8$~V to deplete beneath, but not between, these gates. 
Finally, a Ti/Au gate was placed over the superconducting meander and the tap lines to the device (not shown in Fig.~\ref{fig1}(c) for visibility). The voltage on this gate was set to a large negative voltage, $V_{m}\simeq-6$~V, to ensure that current only travels via the meander and to enhance superconductivity by minimizing the inverse proximity effect from the semiconductor.

\begin{figure}[h]
\centering
\includegraphics[clip, trim=0.2cm 0.2cm 0cm 0.3cm,width=0.48\textwidth]{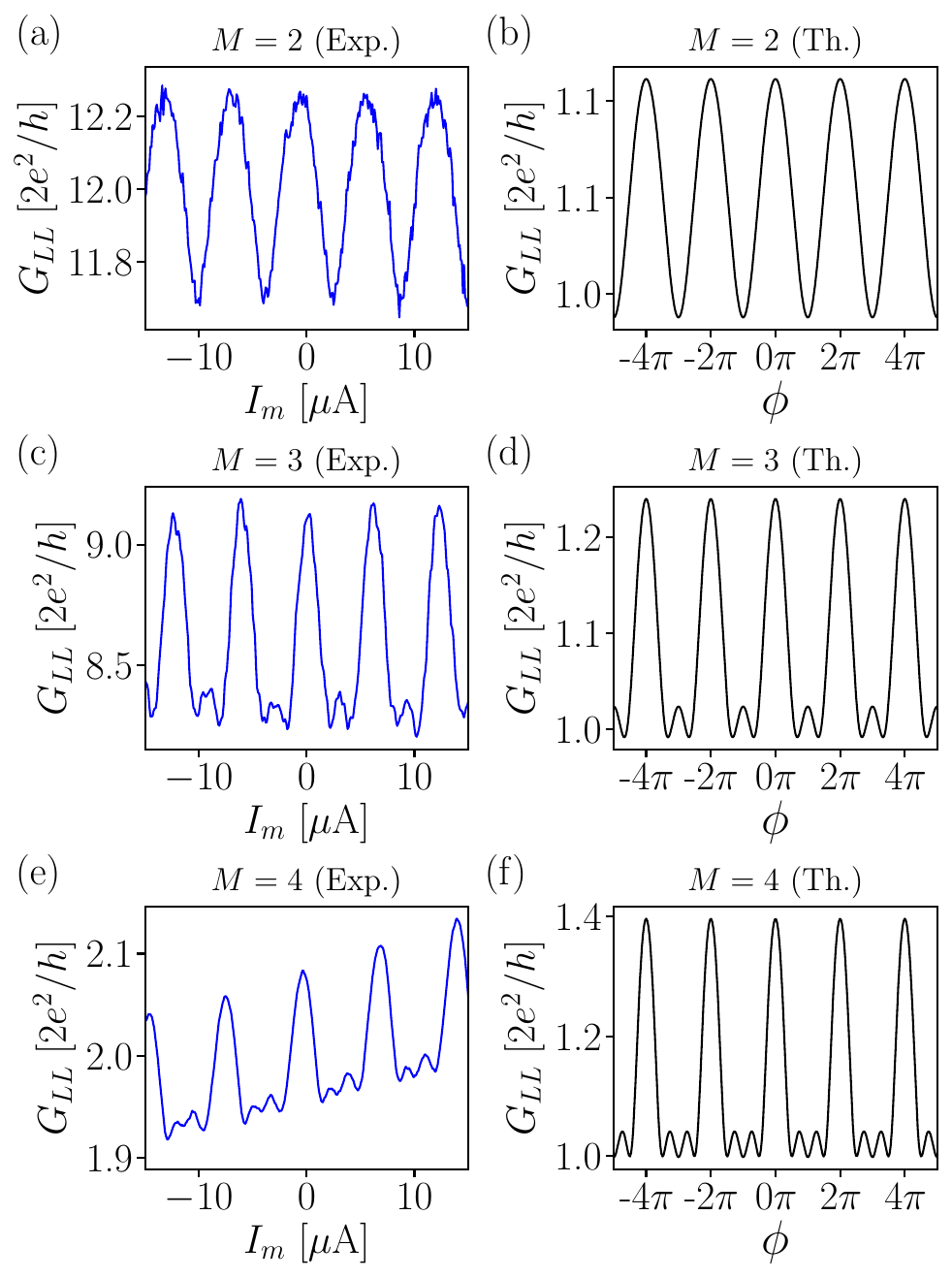}
\caption{Conductance $G_{LL}$, from left lead to ground, in units of $2e^2/h$, for $M=2$, 3, and 4 wire devices, at zero DC bias. (a,c,e) Experimental measurement of $G_{LL}(I_m)$ as a function of current through the meander, $I_{m}$. (b,d,f) Model calculation of $G_{LL}(\phi)$,  where $\phi$ is the phase difference between neighboring wires. Both experiment and model show the number of small maxima changes with the number of wires, $M$.\label{fig2}}
\end{figure} 

Applying a current bias, $I_{m}$, less than the critical current to the superconducting meander line results in a superconducting phase, $\phi$, that advances uniformly along the length of the meander (for uniform cross sectional area) \cite{CurrentControl_1995}, \\
\begin{equation}
\phi(x,I_{m})= a \dfrac{I_{m}}{I_C}\dfrac{x}{\xi}, \label{Eq:Perodicity}
\end{equation}
where $I_C$ is the critical current, $\xi$ is the superconducting coherence length, $x$ is the position along the wire, and $a$ is a coefficient of order 1 (See Supplemental Material (SM) \cite{sup}\nocite{Wafer_Kjaergaard,Tinkham,takane_conductance_1992, danon_nonlocal_2020,SciencePy,Kittel,Manfra.2022}). 

Differential conduction between the left lead and ground, $G_{LL}$, as a function of $I_{m}$, at zero  DC  bias, is shown in Fig.~\ref{fig2}. The periodic conduction pattern results from the increased phase difference between the taps. The period, $T\simeq 7\,\mu$A, was device dependent due to the geometry dependence of $I_C$. Patterns of $G_{LL}$ with zero, one, or two, secondary maxima between primary maxima in Fig.~\ref{fig2} qualitatively resemble the familiar N-slit interference pattern in optics \cite{NSlit_Book}. For optical diffraction from a grating with $N$ slits or reflectors, the intensity pattern at position $x$ from the midpoint of the image is proportional to $[\sin(Nx)/\sin(x)]^2$ for small diffraction angles \cite{NSlit_Book}. We note that this proportionality, without the square, is the same as the form of critical current as a function of phase for JJs in parallel with zero inductance \cite{DeLuca.2015,Courtois.1995}. For the ADG, the local conduction qualitatively follows the relation,
\begin{equation}
    G_{LL}\sim |\sin(M\phi)/\sin(\phi)|, \label{eq:DeLuca}
\end{equation}
for $M$ wires and phase difference  $\phi$ between neighboring wires. While Eq.~(\ref{eq:DeLuca}) yields a qualitative agreement to the measurements performed, a more accurate comparison requires calculating the scattering matrix of the system, as described below and shown in Figs.~\ref{fig2}(b,d,f).
\begin{figure}[h]
\centering
\includegraphics[clip, trim=0cm 0cm 0.5cm 1.25cm, width=0.48\textwidth]{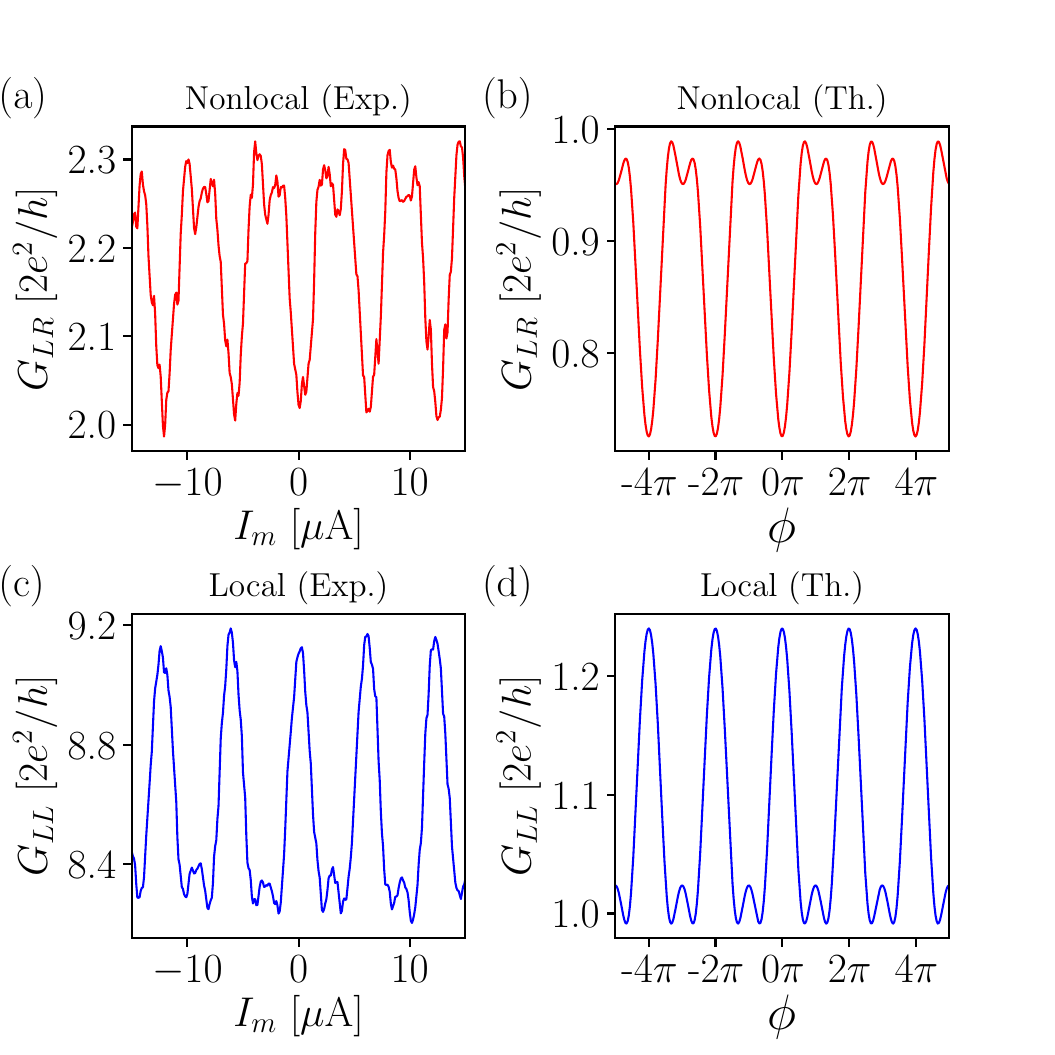}
\caption{Nonlocal conductance, $G_{LR}$,  and local conductance, $G_{LL}$, in units of $2e^2/h$, at zero DC bias. (a,c) Experimental nonlocal and local conductances in the three-wire device as a function of current through the meander, $I_{m}$. (b,d) Model nonlocal and local conductances as a function of phase difference, $\phi$, between neighboring wires. Nonlocal and local signals are inverted, consistent between experiment and model.\label{fig3}}
\end{figure} 

We also measure nonlocal conductances such as $G_{LR}=dI_L/dV_R$, in this case by applying a voltage on the right lead and measuring the current into the left lead. Figures \ref{fig3}(a,c) show the measured nonlocal conductance, $G_{LR}$, and local conductance, $G_{LL}$, for the three-wire device. The observed inversion, where maxima of $G_{LL}$ aligned with minima of $G_{LR}$, is consistent with the model, as seen in Figs.~\ref{fig3}(b,d).

To model the ADG, we calculate the zero-temperature differential conductance matrix, $G_{\alpha\beta}$, from the scattering matrix, $\hat{S}$, which connects  outgoing $(c_{\mathrm{out}})$  incoming $(c_{\mathrm{in}})$ amplitudes of electrons and holes in the leads, $c_{\mathrm{out}} = \hat{S} c_{\mathrm{in}}$. We consider single-channel leads \cite{entin-wohlman_conductance_2008} and find $\hat{S}$ by solving the Bogoliubov–de Gennes (BdG) equations \cite{blonder_transition_1982},
\begin{equation}\label{eq:BdG}
    \mqty( -\frac{\hbar^2}{2m_{\mathrm{eff}}} \frac{\partial^2}{\partial x^2} - \mu(x) & \Tilde{\Delta}(x) \\ \Tilde{\Delta}^*(x) & \frac{\hbar^2}{2m_{\mathrm{eff}}} \frac{\partial^2}{\partial x^2} + \mu(x) ) \bm{\Psi}(x) = \tilde{E} \bm{\Psi}(x),
\end{equation}
 matching wave functions in the normal metal and superconducting regions. In Eq.~(\ref{eq:BdG}),  $\Tilde{\Delta}(x)$ is the proximitized superconducting gap in the semiconductor and $\tilde{E}$ is the renormalized energy of the states (See SM Sec.~\ref{sup:S2} for details \cite{sup}).
This relatively simple model captures the  behavior of the data: Principal maxima are periodic, additional smaller maxima appear with additional wires, and nonlocal versus local signals are inverted. On the other hand, the magnitudes of conductances differ considerably between theory and experiment. This is presumably due to multiple modes between wires and the specifics of the constriction gates, which could be included in a more sophisticated model. By gate-tuning the carrier density between wires, the overall conduction changes without changing the general behavior, shown in Figs.~\ref{fig2} and \ref{fig3}, (Fig.~S3-S6 in SM Sec.~VI \cite{sup}).
\begin{figure}[h] 
\includegraphics[clip, trim=0.2cm 0.2cm 0cm 0.2cm, width=0.48\textwidth]{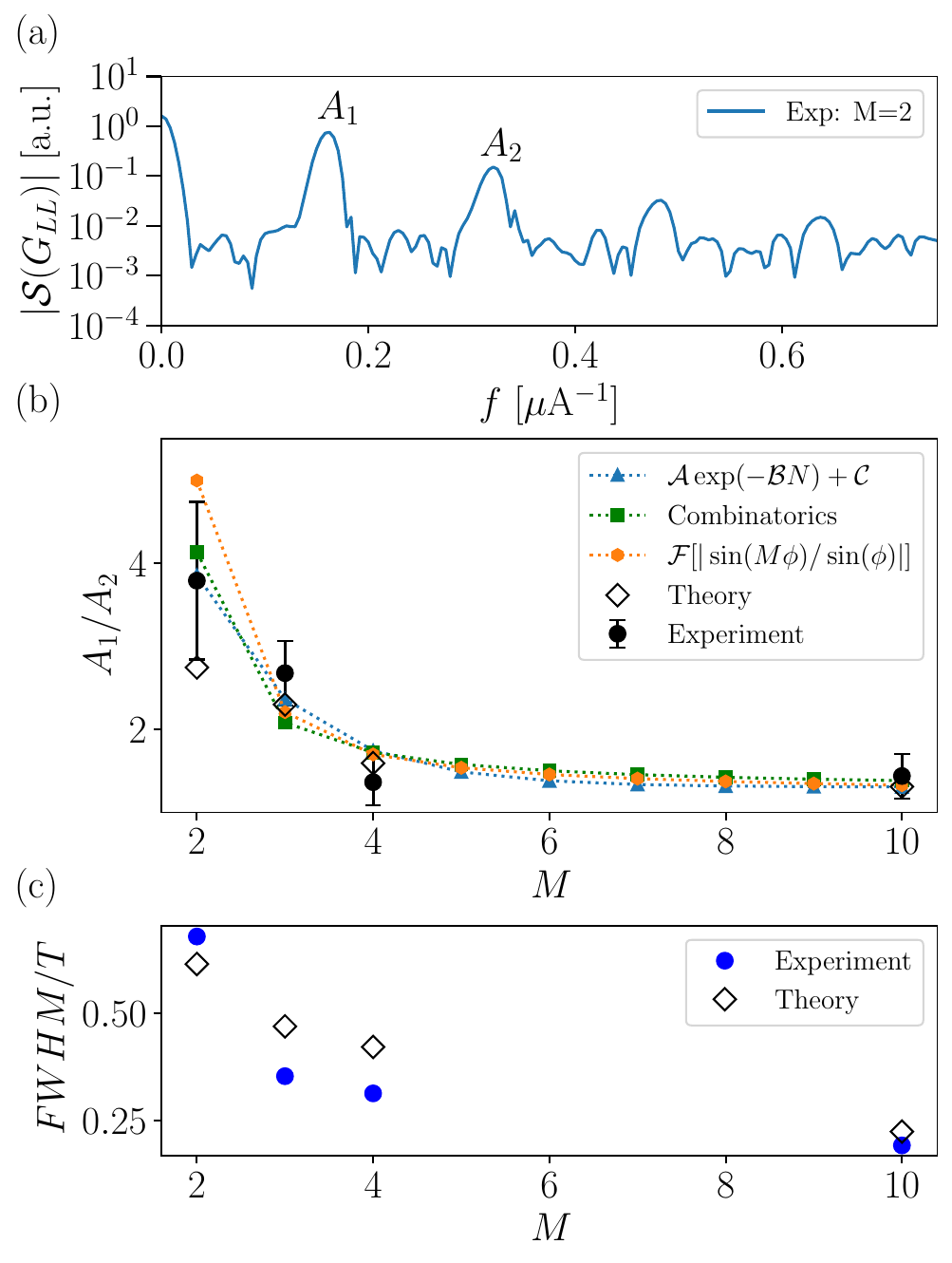}
\caption{(a) Absolute value of the Fourier spectral density, $|\mathcal{S}(G_{LL})|$ for local conductance $G_{LL}(I_m)$ for the $M=2$ device. The first two primary harmonics are labeled $A_1$ and $A_2$. (b) Ratio $A_1/A_2$, as denoted in (a), as a function of the number of wires, $M$. Experimental values (black circles) averaged over gate-controlled JJ density (see text) for  $M=2$, 3, 4, and 10 wires; model values (black diamonds); exponential fit (blue dotted line), with parameters $\mathcal{A}=24.4$, $\mathcal{B}=1.07$, and $\mathcal{C}=1.32$; Fourier transform of  $|\sin(M\phi)/\sin(\phi)|$ (orange circles); and combinatorial analysis (green square), counting harmonics, see Eq.~(S24) \cite{sup}. (c) Full-width at half-max (FWHM) of principal maxima normalized by the period, $T$, of primary maxima. As with optical gratings, the peaks narrow with increasing $M$. \label{fig4}} 
\end{figure}
\begin{figure*}
\centering
\includegraphics[clip, trim=2.3cm 1.2cm 0cm 2.2cm, width=1.05\textwidth]{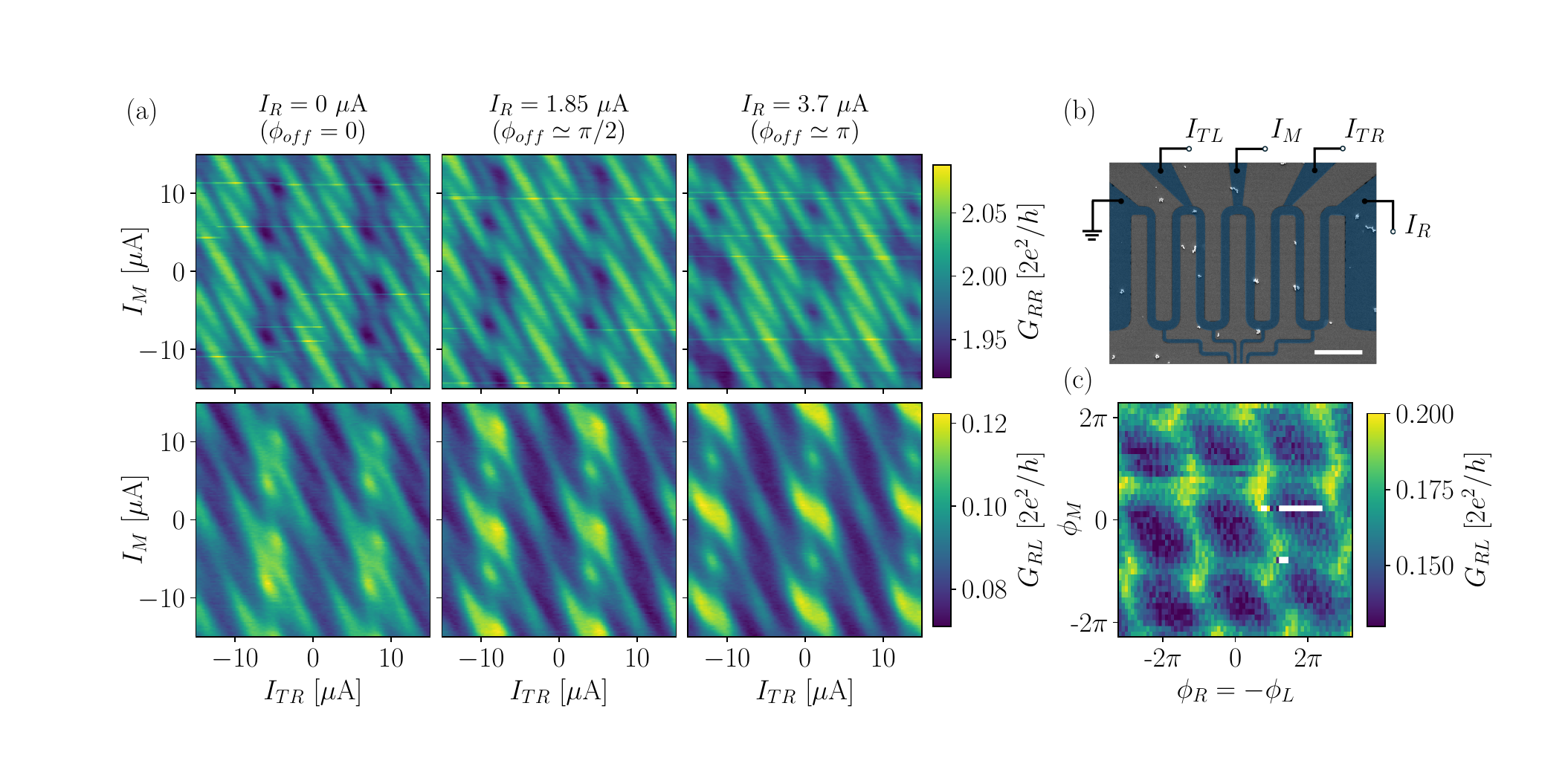}
\caption{(a) Color maps of local and nonlocal conductances, $G_{RR}$ and $G_{RL}$, as a function of tap currents $I_M$ and $I_{TR}$ for $M=4$ device, at zero DC bias. Current $I_R$, applied on the right end of the meander [see (b)] sets a phase-difference offset, $\phi_{\rm off}$, between all neighboring wire pairs.  Three columns (left to right) correspond to $\phi_{\rm off}\sim0$, $\pi/2$, and $\pi$. (b) False color SEM of the meander of the four-wire device, showing current at the top-left, middle, top-right taps, $I_{TL}$, $I_{M}$, $I_{TR}$, and current injected into the meander at the right, $I_{R}$. The scale bar is 5$\mu$m. (c) Nonlocal conductance, $G_{RL}$, as a function of $\phi_R=-\phi_L$ versus $\phi_M$, set using tap currents via Eq.~(\ref{Eq:matrix}).  \label{fig5}}
\end{figure*} 
Spectral features of local conductance are examined in Figs.~\ref{fig4}(a,b), comparing the model and other simple fits to experimental data. 
The Fourier spectrum of local conductance, $|\mathcal{S}[G_{LL}(I_m)]|$, is shown in Fig.~\ref{fig4}(a) (note the logarithmic vertical scale). 
The ratio of the first and second spectral peaks, $A_1/A_2$ [see labels in Fig.~\ref{fig4}(a)], averaged over a range of 0.1~V on the gate between wires (See Fig.~\ref{fig1}(b)), is shown in Fig.~\ref{fig4}(b), along with the model and several other fits. 

As the number of wires, $M$, increases, spectral weight is redistributed to higher frequencies, reflecting narrower Andreev resonances among multiple wires. 
The experimentally observed decrease of $A_1/A_2$ with $M$ [black circles in Fig.~\ref{fig4}(b)] is roughly exponential with an offset of order 1 [blue triangles], and is well approximated by the calculating $A_1/A_2$ from the Fourier transform of $|\sin(M\phi)/\sin(\phi)|$ from Eq.~(\ref{eq:DeLuca}). Besides comparing the fall-off of $A_1/A_2$ to the model (open diamond), we also compare to a simple combinatorial argument of counted the number of higher harmonics from combinations the $M-1$ JJs in the grating, as discussed in SM Sec.~\ref{sup:S4} \cite{sup}). Recalling that for optical gratings increasing the number of slits narrows spectral features, we also examine the full-width at half-max (FWHM) of the principal maxima (for instance, the peaks in Fig.~\ref{fig2}) as a function of $M$. The FWHM, normalized by the period, $T$, is shown in Fig.~\ref{fig4}(c). We observe that the FWHM/$T$ continues to decrease with increasing $M$, even though not all smaller maxima are visible at $M=10$. 

Finally, we demonstrate independent control of phases on individual wires by fabricating additional current-carrying taps off of the meander, along with the phase-biasing taps, as shown in Fig.~\ref{fig5} for the case of $M = 4$. Focusing on the $M=4$ device, with four phase taps and three additional current-carrying taps off of the meander, we define phase differences $\phi_L$ between phase-taps to wires 1 and 2, $\phi_M$ between wires 2 and 3, and $\phi_R$ between wires 3 and 4. Three additional taps along the top of the meander carry currents $I_{TL}$ (top-left), $I_M$ (middle), and $I_{TR}$ (top-right). The right end of the meander carries total current $I_R$, and the left end is grounded. For current taps placed midway between phase taps, phase differences can be related to currents by 
\begin{equation}
    \begin{pmatrix}
        \phi_R\\
        \phi_M\\
        \phi_L
    \end{pmatrix} = \dfrac{a}{I_C}\dfrac{d}{\xi}
    \begin{pmatrix}
        1 & 1/2 & 0 \\
        1 & 1 & 1/2 \\
        1 & 1 & 1
    \end{pmatrix}
    \begin{pmatrix}
        I_R \\ I_{TR} \\ I_M
    \end{pmatrix}, \label{Eq:matrix}
\end{equation}
where $d\simeq24\,\mu$m is the distance between phase taps. The $1/2$ entries in Eq.~(\ref{Eq:matrix}) are a result of the geometry where the additional current taps are placed midway between phase taps. The fourth tap current, $I_{TL}$, is not used here, but can be employed to offset any flux coupling from external fields through loops formed between the meander and the wires.  Figure \ref{fig5}(a) shows local and nonlocal conductance as a function of $I_M$ and $I_{TR}$, for three values of $I_R$ at zero DC bias. 
 Using Eq.~(\ref{Eq:matrix}) allows individual phase differences, $\phi_R$, $\phi_M$, and $\phi_L $ to be set using the current taps. This is demonstrated in Fig.~\ref{fig5}(c), which shows that by setting $\phi_R = -\phi_L$ we observe principal maxima without secondary maxima along two axes, characteristic of control of an effective single junction, consistent with the model (See SM Sec~\ref{sup:S3} \cite{sup}).

In summary, we have introduced and investigated the Andreev diffraction grating (ADG), realized using a gated-patterned superconductor-semiconductor heterostructure. Phases in the ADG were control using a remote current-carrying superconducting meander with phase taps, analogous to screen position in an optical grating. Experimental results in both local and nonlocal conductance show reasonable agreement with a single-channel model of Andreev scattering from multiple phase-biased wires. Adding current-carrying taps to the meander allows control of individual phase differences. Besides exploring the interesting analogy to optical diffraction, with retro- rather than normal reflection, the use of phase- and current-tapped meanders, rather than applied magnetic flux, may have technological relevance when individual phase control is needed or when applying a global magnetic field or local fluxes is not practical, for instance due to cross-talk. 

This research was funded by a research grant (Project 43951) from VILLUM FONDEN, the European Research Council (Grant Agreement No. 856526), and by the DFG Collaborative Research Center (CRC) 183 Project No. 277101999.
\begin{center}
\begin{tikzpicture}
    \draw[thick, line width=0.5pt] (0,0) -- (0.75,0);
    \draw[thick, line width=0.8pt] (0.75,0) -- (1.5,0);
    \draw[thick, line width=1.1pt] (1.5,0) -- (2.25,0);
    \draw[thick, line width=1.4pt] (2.25,0) -- (3.75,0);
    \draw[thick, line width=1.1pt] (3.75,0) -- (4.5,0);
    \draw[thick, line width=0.8pt] (4.5,0) -- (5.25,0);
    \draw[thick, line width=0.5pt] (5.25,0) -- (6,0);
\end{tikzpicture}
\end{center}
\bibliographystyle{apsrev4-2}
\bibliography{main_bib}
\onecolumn
\renewcommand{\thesection}{\Roman{section}}
\renewcommand{\thefigure}{S\arabic{figure}}
\setcounter{section}{0}
\setcounter{figure}{0}
\setcounter{table}{0}
\setcounter{page}{1}
\setcounter{equation}{0}
\renewcommand{\theequation}{S\arabic{equation}}
\begin{center}
  \textbf{\large Supplementary Material for Phase-Biased Andreev Diffraction Grating}\\[.2cm]
  Magnus~R.~Lykkegaard,$^{1}$ Anders~Enevold~Dahl,$^{1}$ Karsten~Flensberg,$^{1}$ \\Tyler~Lindemann,$^{2,3}$ Michael~J.~Manfra,$^{2,3,4,5}$ and Charles~M.~Marcus$^{1,6,7}$\\[.1cm]
  {\itshape ${}^1$Center for Quantum Devices, Niels Bohr Institute, \\ University of Copenhagen, DK-2100 Copenhagen, Denmark\\
  ${}^2$Department of Physics and Astronomy, Purdue University, West Lafayette, Indiana 47907, USA\\
  ${}^3$Birck Nanotechnology Center, Purdue University, West Lafayette, Indiana 47907, USA\\}
  ${}^4$School of Electrical and Computer Engineering,\\ Purdue University, West Lafayette, Indiana 47907, USA\\
  ${}^5$School of Materials Engineering, Purdue University, West Lafayette, Indiana 47907, USA\\
  ${}^6$Department of Physics, University of Washington, Seattle, Washington 98195, USA\\
  ${}^7$Materials Science and Engineering, University of Washington, Seattle, Washington 98195, USA\\
(Dated: \today)\\[1cm]
\end{center}

\section{Method/Setup/Material \label{sup:S0}} 
The four measured devices, with two, three, four, and ten Andreev wires, were fabricated on epitaxially matched InAs/Al heterostructures. In the heterostructures, the InAs layer forms a shallow two-dimensional electron gas (2DEG) with high transparency between the quantum well and the superconducting Al layer \cite{Wafer_Kjaergaard}. The heterostructure stack consists of a InP wafer, followed by a graded In$_{\rm 1-x}$Al$_{\rm x}$As buffer with $x$ running from 0.48 at the InP wafer interface to 0.19 at the bottom barrier.  The quantum well consists of In$_{0.75}$Ga$_{0.25}$As barriers, and a 7 nm InAs quantum well. The top barrier layer is 10 nm thick and a 5 nm epitaxial Al layer is grown at low temperature.  The superconducting regions---superconducting wires, pick-up lines, leads, and meander--- were formed by e-beam lithography and wet-etching using Transene D, a commercial Al etchant. The Ti/Au gates are deposited after a blanket atomic layer deposition of HfO$_2$ by electron beam lithography and evaporation step. The individual devices are also electrically isolated with a deep mesa etch using a solution of H$_2$O:C$_6$H$_8$O$_7$:H$_3$PO$_4$:H$_2$O$_2$ in ratio of 220:55:3:3.

The connections along the meander to the superconducting wires, are separated by several coherence lengths of the superconductor, $d\sim 15\, \xi$, using the intrinsic Pippard coherence length of Al, $\xi = 1.6$~$\mu$m \cite{Kittel}. This allows the phase difference across the JJs to wind by several $\pi$ while remaining well below the critical current of the meander. Though we expect the actual coherence length of the aluminum meander to be smaller, this design choice sets a lower bound on the number of phase windings we can produce. More likely, we are in the dirty/thin superconductor limit, with a diffusive coherence length $\xi_d=\sqrt{l_ev_F\hbar/(2\Delta)}\simeq490$~nm, using a scattering length $l_e=\hbar\mu\sqrt{2\pi n}/e\simeq220$~nm. We use the mobility, fermi velocity, density and effective mass, $\mu,v_F,n,m^*$ respectively, found in Ref. \cite{Manfra.2022}. The critical current of the meander in the devices that were investigated (with a width $W\simeq 1$~$\mu$m and thickness of $h\simeq 5$~nm) is on the order of $I_C\simeq 90$~$\mu$A, which is congruent with other experiments done on equivalent heterostructures. For the two wire device, with a 48~$\mu$m long and 1~$\mu m$ wide wire, we find a normal state resistance, $R_N\simeq700\Omega$ above the critical current, or normal state resistivity $\rho_N=14.6$~$\Omega$, which corresponds to a scattering length $l_e=225$~nm.

We can also compare the geometric inductance of the square loop with the Josephson inductance of the individual JJs. We find a geometric inductance $L_{G}\simeq17$~pH. We approximate a normal resitance of the junctions $R_N=100~\text{nm}/(1~\mu\text{m} \cdot \sigma)= 100~\text{nm}/(1~\mu\text{m} \cdot n\cdot e \cdot\mu)\simeq50 ~\Omega$, and from that a critical current $I_C=\pi\Delta/(2eR_N)=5.7~\mu$A. From this we find at $\phi=0$ the Josephson inductance is $L_J(\phi=0)=\hbar/(2 e I_C)\simeq 57.7$~pH.

\begin{figure}[h!]
\centering
\includegraphics[width=0.75\textwidth]{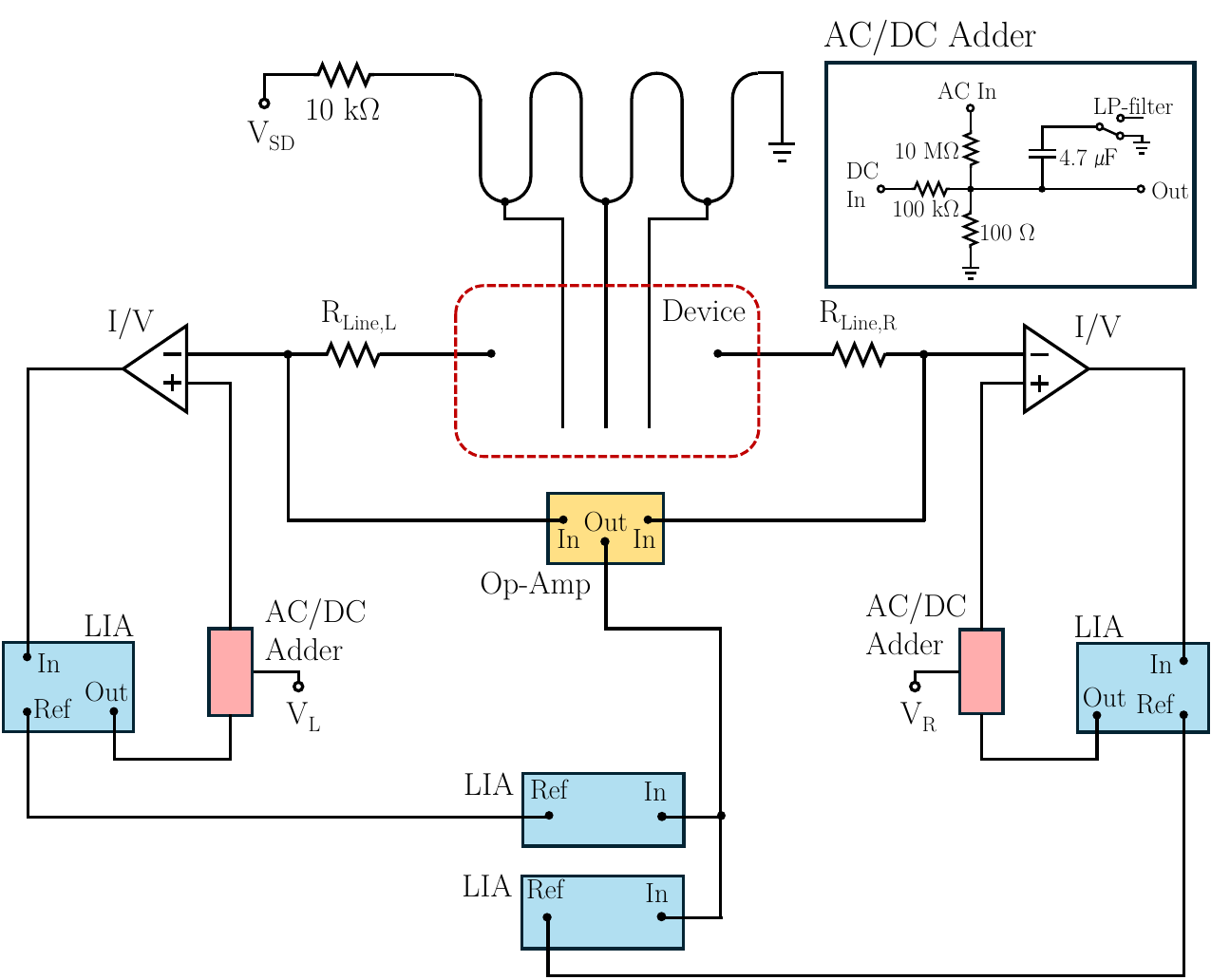}
\caption{The circuit of a three-wire device. The meander is current biased, $I_m$, by attaching a 10~k$\Omega$ resistor after a voltage source, and varying $V_{SD}$. We can measure this response in the current and the voltage. The current is converted to a voltage and amplified through a Basel I/V converter (labeled I/V in the figure). The voltage across the grating is amplified using a differential voltage amplifier (labeled Op-Amp). Two AC signals at frequencies $f_{L}\simeq89$~Hz and $f_{R}\simeq163$~Hz are supplied in addition to an offset DC voltage, $V_{L}$ and $V_{R}$, through an AC/DC adder (which circuitry is shown in the inset). The offset voltage is applied to align the potentials within the device, so no DC current flows. We measured a line resistance $R_{line,i}=1$~k$\Omega$, which includes the resistance of the attenuators and RC filters connected to the DC lines within the dilution refrigerator. \label{figsup:6}}
\end{figure}

Transport measurements were performed using standard AC lock-in methods in a three-terminal setup, in a dilution refrigerator with a mixing-chamber temperature of $20$~mK. The devices has two normal leads on either side of the grated superconductor, labeled $R$ and $L$, which is AC voltage biased with frequencies $f_{L/R}$, and a superconducting lead, labeled $G$, that is grounded to the meander. For each device the optimal $f_{L/R}$, were found by measuring the signal-to-noise ratio at different frequencies. For most of the measurements presented, $f_{L/R} \simeq 89/163$~Hz. For the experiment presented in Fig.~5 of the paper, we found that it had shifted to lower frequencies, $f_{L/R} \simeq 37/107$~Hz, which produced a optimal signal-to-noise ratio. Simultaneously we also optimized the AC excitation voltage applied to the ohmic leads, which we kept $\leq 8$~$\mu$V. Transport is measured from the normal leads into the superconducting ground, $G_{RR}$ and $G_{LL}$ and between normal leads, $G_{RL}$ and $G_{LR}$. A detailed sketch of the circuit for the three wire device is shown in Fig.~\ref{figsup:6}. Using the Basel SP 983 I/V converter we are able to both measure and apply an AC or DC voltage on the same line.

\section{Phase gradient derivation \label{sup:S1}}
Here we explain Eq.~(1) of the main text. Ginzburg-Landau (GL) theory relates the supercurrent in the wire, $I_m$, and phase gradient, $\nabla\phi$, in the absence of an applied magnetic field as \cite{Tinkham},
\begin{equation}
    I_m=\frac{An_s e \hbar}{m_e} \grad \phi,
\end{equation}
where $n_s$ is the superfluid density and $A$ is the cross-sectional area of the superconducting wire. Note that we use GL to describe the phase winding within the superconducting meandering wire, and assume that the proximitized semiconductor acquires the same phase as the current biased wire. This equation hold for thin superconducting wires where the current is uniformly distributed. By integrating the gradient along the position of the wire, a coordinate $x$, we are left with,
\begin{equation}
    \phi(x,I_{m})= a \frac{I_{m}}{I_C}\frac{x}{\xi},
\end{equation}
where $\phi(x=0)=0$,  $I_C\xi=An_se\hbar/m^*$, and $a$ is a coefficient of order 1, which in general will be dependent on material and geometry of the device. Taking $I_C=4eAn_s/3(2\alpha(T)/3m^*)^{1/2}$ and $\xi^2=\hbar^2/2m^*\alpha(T)$, as the critical current and coherence length from GL theory yields, where $\alpha(T)$ is a parameter of the GL equation,
\begin{equation}
    I_{C,GL}\xi_{GL}= \dfrac{4}{3^{3/2}}Aen_s\hbar/m^*,
\end{equation}
where $m^*$ is the effective mass (Sec. 4.4 of \cite{Tinkham}).
\section{Theory derivations of scattering matrix \label{sup:S2}}
In the main text we show simulations of the device. Here we show the theoretical derivation of the scattering matrix approach to find the zero-temperature differential conductance. The scattering matrix, $\hat{S}$, links the incoming amplitudes of electrons and holes to the outgoing amplitudes,

\begin{align}
    c_{\mathrm{out}} = \hat{S} c_{\mathrm{in}}, \label{eq:scattering_req}
\end{align}
where we study the single channel regime,
\begin{align}
    c_{\mathrm{in}} = \qty( C_e^+(L),  C_h^-(L),  C_e^-(R), C_h^+(R) )^T, \quad c_{\mathrm{out}} = \qty( C_e^-(L), C_h^+(L), C_e^+(R),  C_h^-(R) )^T,
\end{align}
where $\hat{S}$ takes the form

\begin{align}
    \hat{S} = \mqty( \bm{r_{LL}} & \bm{t_{LR}} \\
                \bm{t_{RL}} & \bm{r_{RR}} ) = 
                \mqty(
                    S_{LL}^{ee} & S_{LL}^{eh} & S_{LR}^{ee} & S_{LR}^{eh} \\
                    S_{LL}^{he} & S_{LL}^{hh} & S_{LR}^{he} & S_{LR}^{hh} \\
                    S_{RL}^{ee} & S_{RL}^{eh} & S_{RR}^{ee} & S_{RR}^{eh} \\
                    S_{RL}^{he} & S_{RL}^{hh} & S_{RR}^{he} & S_{RR}^{hh}). \label{eq:scattering_form}
\end{align}
In order to find the scattering matrix for a NSN junction we solve the Bogoliubov–de Gennes (BdG) equations in both a normal and superconducting region.

\begin{equation}
    \mqty( -\frac{\hbar^2}{2m_{\mathrm{eff}}} \frac{\partial^2}{\partial x^2} - \mu(x) & \Tilde{\Delta}(x) \\ \Tilde{\Delta}^*(x) & \frac{\hbar^2}{2m_{\mathrm{eff}}} \frac{\partial^2}{\partial x^2} + \mu(x) ) \bm{\Psi}(x) = \tilde{E} \bm{\Psi}(x),
\end{equation}
where $\Tilde{\Delta}(x)$ is the proximitized superconducting gap of the semiconductor and $\tilde{E}$ is the renormalized energy of the states in the superconducting region. In order to find the proximitized gap and the renormalized energy we use the retarded Green's function,

\begin{align}
    G^R(\omega, k) = \frac{1}{\omega(1 + \frac{\Gamma_0}{\sqrt{\Delta_0^2 - \omega^2}}) \tau_0 - \xi_k \tau_0 + \frac{\Gamma_0 \Delta_0}{{\sqrt{\Delta_0^2 - \omega^2}}} \tau_1 },
\end{align}
which tell us $\tilde{\Delta} = \frac{\Gamma_0 \Delta_0}{\sqrt{\Delta_0^2 - E^2}}$ and $\tilde{E} = E\qty(1 + \frac{\Gamma_0}{\sqrt{\Delta_0^2 - E^2}})$. The wave functions in the superconducting region is then after solving the BdG equations,
\begin{align}
    \psi_1^\pm(x) &= \mqty( \mathrm{e}^{\eta_1/2} \\ \mathrm{e}^{\eta_2/2} ) \qty( \qty(\tilde{E} \tilde{E}' - \tilde{\Delta} \tilde{\Delta}') )^{1/2} \qty(2q_1)^{-1/2} \qty(\tilde{E}^2 - \tilde{\Delta}^2)^{-1/4} \mathrm{e}^{\pm i q_1 x} \\
    \psi_2^\pm(x) &= \mqty( \mathrm{e}^{\eta_1/2} \\ \mathrm{e}^{\eta_2/2} ) \qty( \qty(\tilde{E} \tilde{E}' - \tilde{\Delta} \tilde{\Delta}') )^{1/2} \qty(2q_2)^{-1/2} \qty(\tilde{E}^2 - \tilde{\Delta}^2)^{-1/4} \mathrm{e}^{\pm i q_2 x}
\end{align}
where

\begin{align}
    \eta_{1,2} &=  \phi \pm \arccos\qty(\tilde{E}/\tilde{\Delta}),\\
    \tilde{E}' &= \frac{\partial \tilde{E}}{\partial E} = 1 + \frac{\Gamma_0 E^2}{\qty(\Delta_0^2 - E^2)^{3/2}} + \frac{\Gamma_0}{\sqrt{\Delta_0^2 - E^2}}, \\
    \tilde{\Delta}' &=  \frac{\partial \tilde{\Delta}}{\partial E} = \frac{\Gamma_0 \Delta_0 E}{\qty( \Delta_0^2 - E^2 )^{3/2}},
\end{align}
and the wave numbers are,

\begin{align}
    q^2_{1,2} = \frac{2m}{\hbar^2} \qty(\mu_S \pm \sqrt{\tilde{E}^2 - \tilde{\Delta}^2}).\label{eq:wavenumbers}
\end{align} 
In the normal regions we solve the BdG equations with no superconducting gap, and the solutions are,

\begin{align}
    \psi_e^\pm(x) = \mqty(1 \\ 0) \qty(k_e)^{-1/2} \mathrm{e}^{\pm ik_e x}, \quad \psi_h^\pm(x) = \mqty(0 \\ 1) \qty(k_h)^{-1/2} \mathrm{e}^{\pm ik_h x},
\end{align}
with
\begin{align}
    k^2_{e,h} = \frac{2m}{\hbar^2} \qty( \mu_N \pm E ).\label{eq:wavenumbersk}
\end{align} 

In order to find $\hat{S}$ we construct the wave functions in both the superconducting and normal regions to be,

\begin{align}
    \Psi_N(x) = a_e^+ \psi_e^+(x) + a_e^- \psi_e^-(x) + a_h^+ \psi_h^+(x) + a_h^- \psi_h^-, \\ \nonumber\\
    \Psi_S(x) = b_1^+ \psi_1^+(x) + b_1^- \psi_1^-(x) + b_2^+ \psi_2^+(x) + b_2^- \psi_2^-.  
\end{align}
We find the scattering matrix $\hat{S}_{NSN}$ which is for a normal-super-normal junction and the scattering matrix $\hat{S}_{NNN}$ corresponding to a normal-normal-normal junction. Using these scattering matrices we can concatenate them in order to find the full scattering matrix for the device we wish to simulate.

\subsection{S matrix concatenation}

We can concatenate the NSN scattering matrix together with the NNN matrix, in order to make setups with more superconducting fingers.
If we have two scattering matrices

\begin{align}
     \hat{S}_1 = \begin{pmatrix} \mathbf{r_{L1}} & \mathbf{t_{LR1}} \\
    \mathbf{t_{RL1}} & \mathbf{r_{R1}} \end{pmatrix}, \quad \hat{S}_2 = \begin{pmatrix} \mathbf{r_{L2}} & \mathbf{t_{LR2}} \\
    \mathbf{t_{RL2}} & \mathbf{r_{R2}} \end{pmatrix}, 
\end{align}

and we wish to find the full scattering matrix, $\hat{S}_{12}$, where both regions are included,

\begin{align}
    \hat{S}_{12} = \begin{pmatrix} \mathbf{r_{L12}} & \mathbf{t_{LR12}} \\
    \mathbf{t_{RL12}} & \mathbf{r_{R12}} \end{pmatrix},
\end{align}

then we get

\begin{align}
    \mathbf{r_{L12}} &= \mathbf{r_{L1}} + \mathbf{t_{LR1}r_{L2}} ( \mathds{1} - \mathbf{r_{R1}r_{L2}} )^{-1} \mathbf{t_{RL1}}, \nonumber\\
    \mathbf{r_{R12}} &= \mathbf{r_{R2}} + \mathbf{t_{RL2}}( \mathds{1} - \mathbf{r_{R1}r_{L2}} )^{-1}\mathbf{r_{R1}}  \mathbf{t_{LR2}}, \nonumber\\
    \mathbf{t_{RL12}} &= \mathbf{t_{RL2}} (\mathds{1} - \mathbf{r_{R1}r_{L2}})^{-1} \mathbf{t_{RL1}} ,\nonumber\\
    \mathbf{t_{LR12}} &=  \mathbf{t_{LR1}} (\mathds{1} + \mathbf{r_{L2}} (\mathds{1} - \mathbf{r_{R1} r_{L2}})^{-1} \mathbf{r_{R1}}) \mathbf{t_{LR2}},
\end{align}
after taking all reflections and transmissions into account.

\subsection{Conductance from S matrix}
We can find the zero-temperature differential conductance from scattering matrix as in \cite{takane_conductance_1992, danon_nonlocal_2020}. For a single channel scattering matrix at zero temperature with form of Eq.~(\ref{eq:scattering_form}) the differential conductance matrix is given by,

\begin{align}
    \mathbf{G} = \begin{pmatrix}
        G_{LL} & G_{LR} \\
        G_{RL} & G_{RR}
    \end{pmatrix} = \frac{2e^2}{h} \begin{pmatrix}
        1 - |S_{LL}^{ee}|^2 + 
        |S_{LL}^{he}|^2 &
         - (|S_{LR}^{ee}|^2 - |S_{LR}^{he}|^2 ) \\
         - (|S_{RL}^{ee}|^2 - |S_{RL}^{he}|^2 ) & 1 - |S_{RR}^{ee}|^2 + 
        |S_{RR}^{he}|^2 
    \end{pmatrix}.
\end{align}

\subsection{Parameters used in simulations}
We have used the following parameters in the simulations. For all simulations we have used,

\begin{align}
    L_N = 100 \text{ nm}, \quad
    L_S = 100 \text{ nm}, \quad
    m_\mathrm{eff} = 0.026 m_e, \quad
     \Delta_0 = 0.2 \text{ meV},\nonumber\\
     \Gamma = \Delta_0, \quad \mu_S = 1.94 \text{ meV}, \quad \mu_N = 1.36 \text{ meV}, \quad \delta\phi = \frac{\pi}{6},
\end{align}
where $\delta\phi$ is an extra phase that accumulates because of a variation in the widths in the meander between where S wires tap onto the S meander. This is only relevant for the three and four wire device. The two chemical potentials corresponds to a transparency, $\mathcal{T} = \left| t \right|^2 = \frac{4 \sqrt{\mu_S \mu_N}}{\left( \sqrt{\mu_S} + \sqrt{\mu_N} \right)^2} = 0.992$. For perfect transparency, the local conductance of the two-wire device resembles a sinusoidal function. Lower transparency results in the local conductance sharpening the cusps and resembles Eq.~(2) more.

\section{Simulation of Fig.~5(c) in main text\label{sup:S3}}
\begin{figure}[h]
    \centering
    \includegraphics[width=0.95\textwidth]{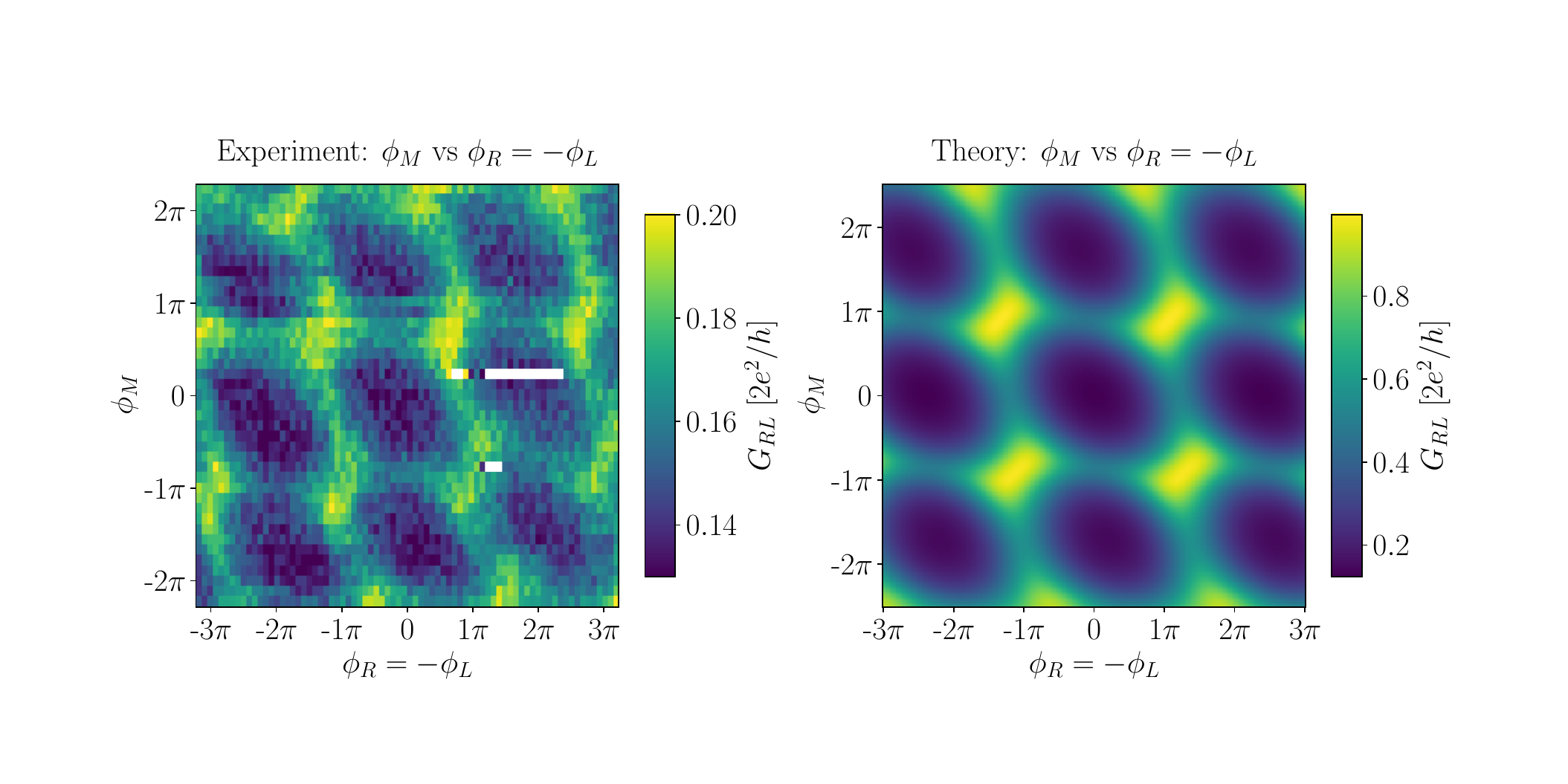}
    \caption{Left: Nonlocal conductance, $G_{RL}$, of the four wire experiment as a function of $\phi_R=-\phi_L$ and $\phi_M$. Right: Simulation of the four wire experiment corresponding to Fig.~5(c) of the main text, where the currents are converted to phase difference between wires}
    \label{fig:sim5c}
\end{figure}

Fig.~\ref{fig:sim5c} shows a comparison of experimental and theoretical nonlocal conductance as a function of $\phi_R=-\phi_L$ and $\phi_M$. The theory follow the procedure discussed in Sec.~\ref{sup:S2}. We observe that the simulation is in agreement with what we measure in the experiment. The white areas in the experimental data is missing data points from the slice of the data cube with axes $(\phi_R,\phi_{M},\phi_{L})$.

\section{Suppression of $A_1/A_2$ \label{sup:S4}}
In the main text we observe that the ratio of coefficients of the absolute value of the spectral density in the FFT, $A_1/A_2$, is suppressed as the number of wires is increased. We can predict this suppression using combinatorics. In our devices JJs formed by neighbouring S wires contribute the base harmonic, next-nearest neighbouring wires contribute 1st harmonic and so on. Furthermore each JJ (either nearest, next-nearest neighbour or higher) can also support higher harmonics originating from multiple Andreev scattering processes.

We solve the problem, by calculating the multiplicity, $\Omega_{k,M}$, of each type of neighbours/higher harmonics: Given $M$ wires how many $k$-type neighbouring wires is there, $\Omega_{k,M}$? Here $k$ is the number of wires between two wires. As each junction can host higher harmonics, we account for this by defining that any $k^{th}$-neighbour is also $m(k+1)$-type neighbours, where $m$ is a positive integer. We also add that that due to the induced coherence length in the JJ, the fourier coefficient is exponentially suppressed by a factor of $\exp(-nd/\xi^*)$, where $n=k+1$ is the n'th order harmonic, $d$ is the relevant size of the JJ (the length in our case), and $\xi^*$ is the induced coherence length of the JJ. That is we predict the strength of the n'th harmonic to be,
\begin{equation}
  A_n(M)=\dfrac{\exp(-nd/\xi^*)\cdot\Omega_{n,M}}{\sum_{l\neq n}\exp(-ld/\xi^*)\cdot\Omega_{l,M}}, \label{eq:combi}
\end{equation}
where we have normalized $\Omega_{n,M}$ to extract the relevance ratio: The ratio between the n$^{th}$ harmonic and sum of all other harmonics, $\sum_{l\neq n}$. In the case of the Andreev diffraction grating of 2, 3, 4 and 10 wires, we compare the measured ratio with that obtained from combinatorics with the fit $d/\xi^*=0.85$ in Fig.~4(b) of the paper. For comparison,  we have showed in the main text the values achieved from a FFT of simple model $\mathcal{F}\{|\sin(M\phi)/\sin(\phi)\}$ and a fit of the ratio to an exponential suppression of the form 
\begin{equation}
\dfrac{A_1}{A_2} = \mathcal{A}\exp(-\mathcal{B}M)+\mathcal{C},
\end{equation}
which gives parameters $\mathcal{A}=24.42$, $\mathcal{B}=1.07$, and $\mathcal{C}=1.32$. Both fits follow the experimentally observed suppression of the ratio. Furthermore, we note that the ratio as the number of wires , $M$, increases reaches an asymptote, in all the cases. For $M\rightarrow\infty$ we see that the combinatorial model approaches a ratio of 1.26, the exponential approaches 1.32 and the FFT of the simple model approaches 1. It is worth to note that the simple model, has no inclusion of an exponential suppression of harmonics due to the sizes of the JJs.

\section{Gate dependence and FFT\label{sup:S5} }
\begin{figure} 
\centering
\includegraphics[width=\textwidth]{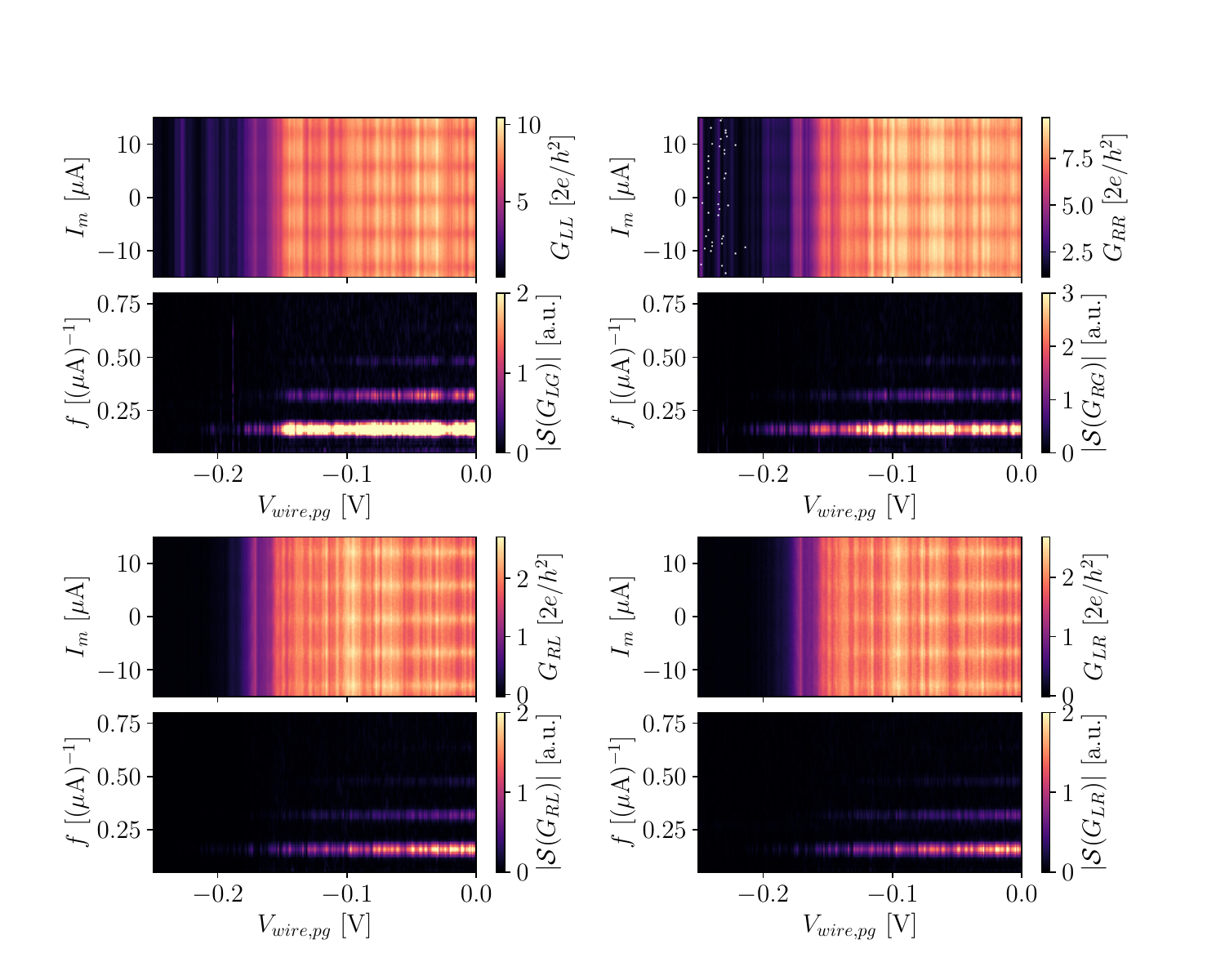}
\caption{Two-wire device: Local conductance, $G_{LL/RR}$, and nonlocal conductance, $G_{LR/RL}$, as a function of the current through the meander, $I_{m}$, and voltage, $V_{wire,pg}$, on the gate between the wires. The lower part of each panel show the absolute value of the spectral density, $|S|$, as a function of frequency, $f$, associated with $I_{m}$.\label{figsup:1}}
\end{figure}

\begin{figure} 
\centering
\includegraphics[width=\textwidth]{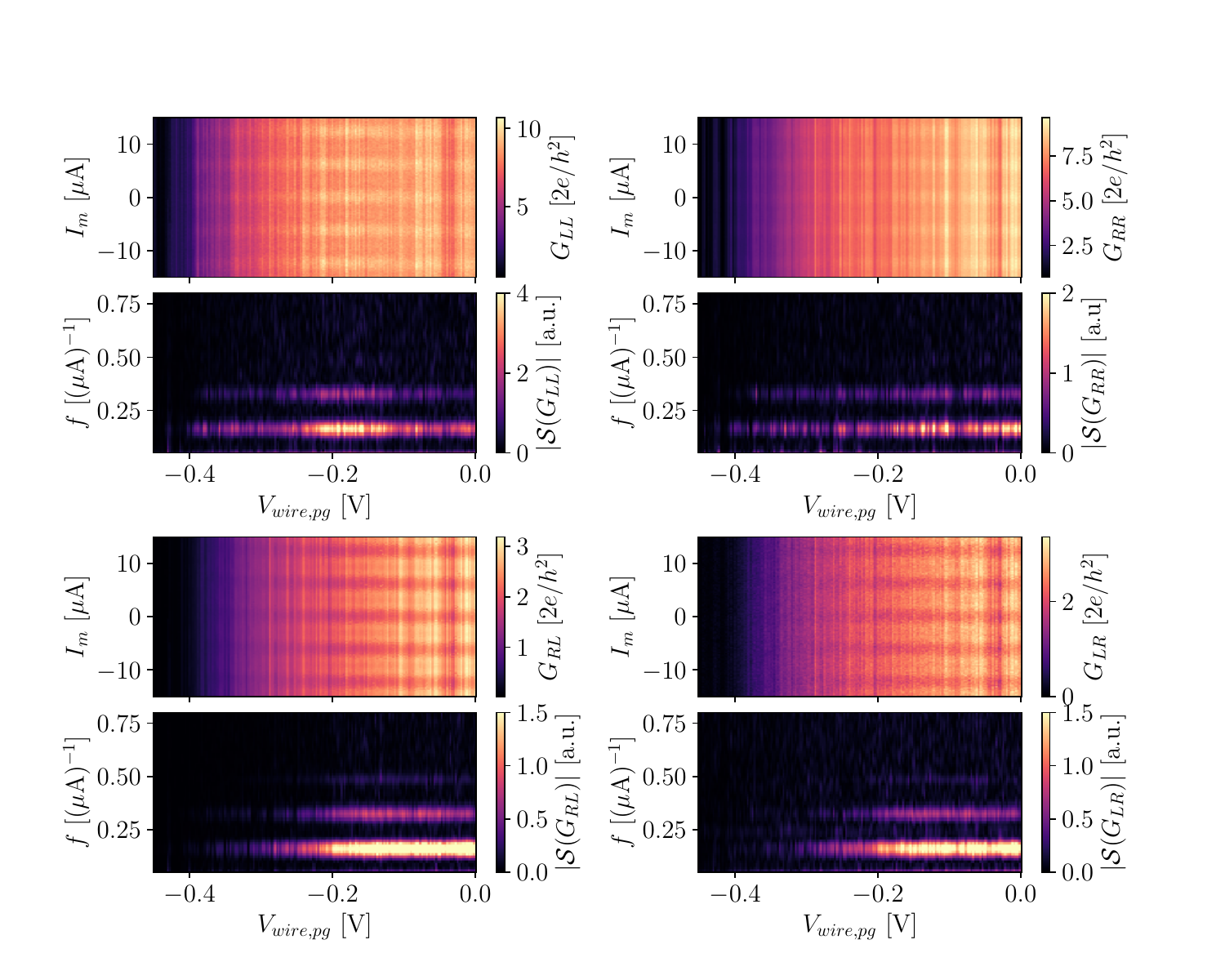}
\caption{Three-wire device: Local conductance, $G_{LL/RR}$, and nonlocal conductance, $G_{LR/RL}$, as a function of the current through the meander, $I_{m}$, and voltage, $V_{wire,pg}$, on the gate between the wires. The lower part of each panel show the  absolute value of the spectral density, $|S|$, as a function of frequency, $f$, associated with $I_{m}$.\label{figsup:2}}
\end{figure}

\begin{figure} 
\centering
\includegraphics[width=\textwidth]{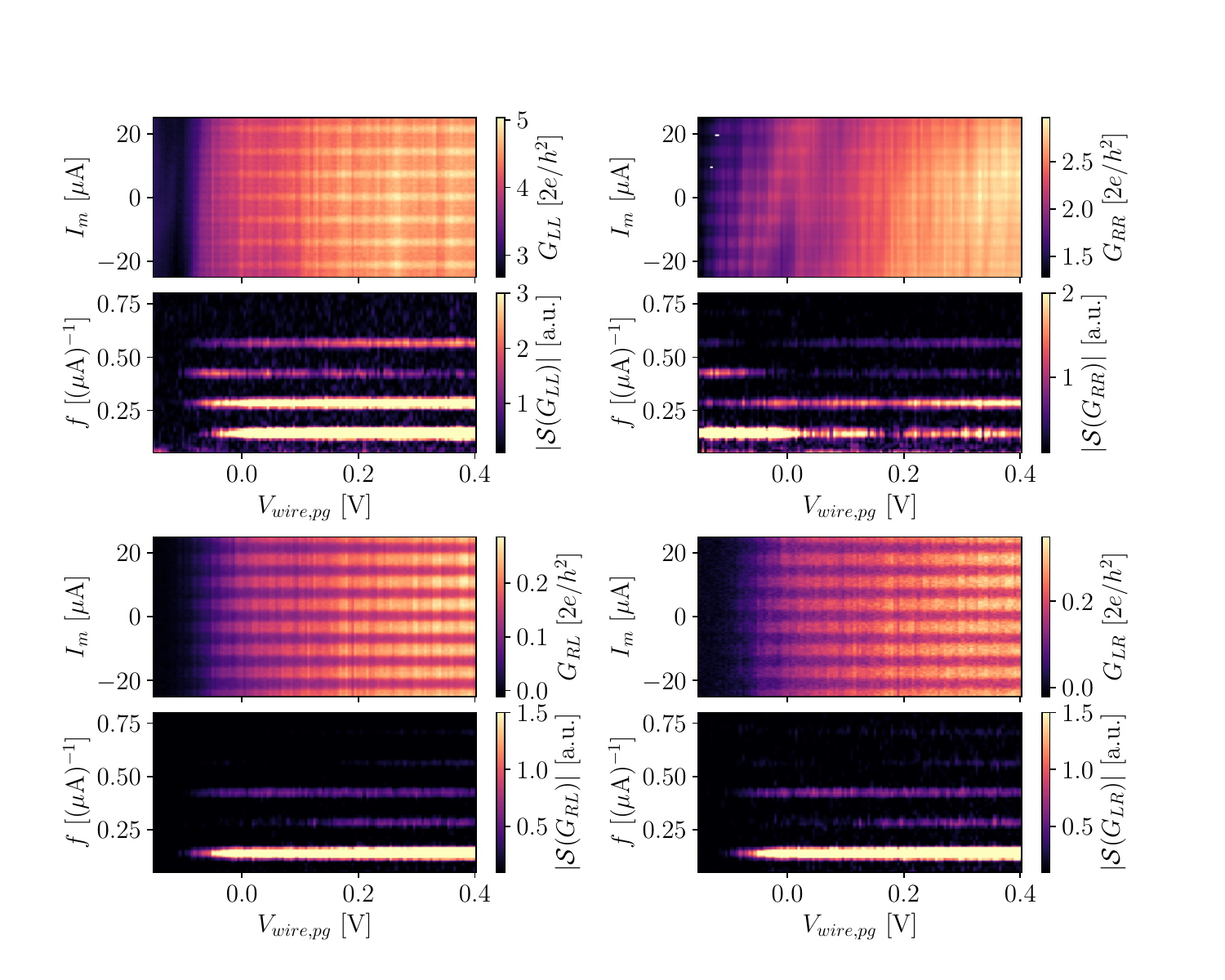}
\caption{Four-wire device: Local conductance, $G_{LL/RR}$, and nonlocal conductance, $G_{LR/RL}$, as a function of the current through the meander, $I_{m}$, and voltage, $V_{wire,pg}$, on the gate between the wires. The lower part of each panel show the absolute value of the spectral density, $|S|$, as a function of frequency, $f$, associated with $I_{m}$. \label{figsup:3}}
\end{figure}

\begin{figure} 
\centering
\includegraphics[width=\textwidth]{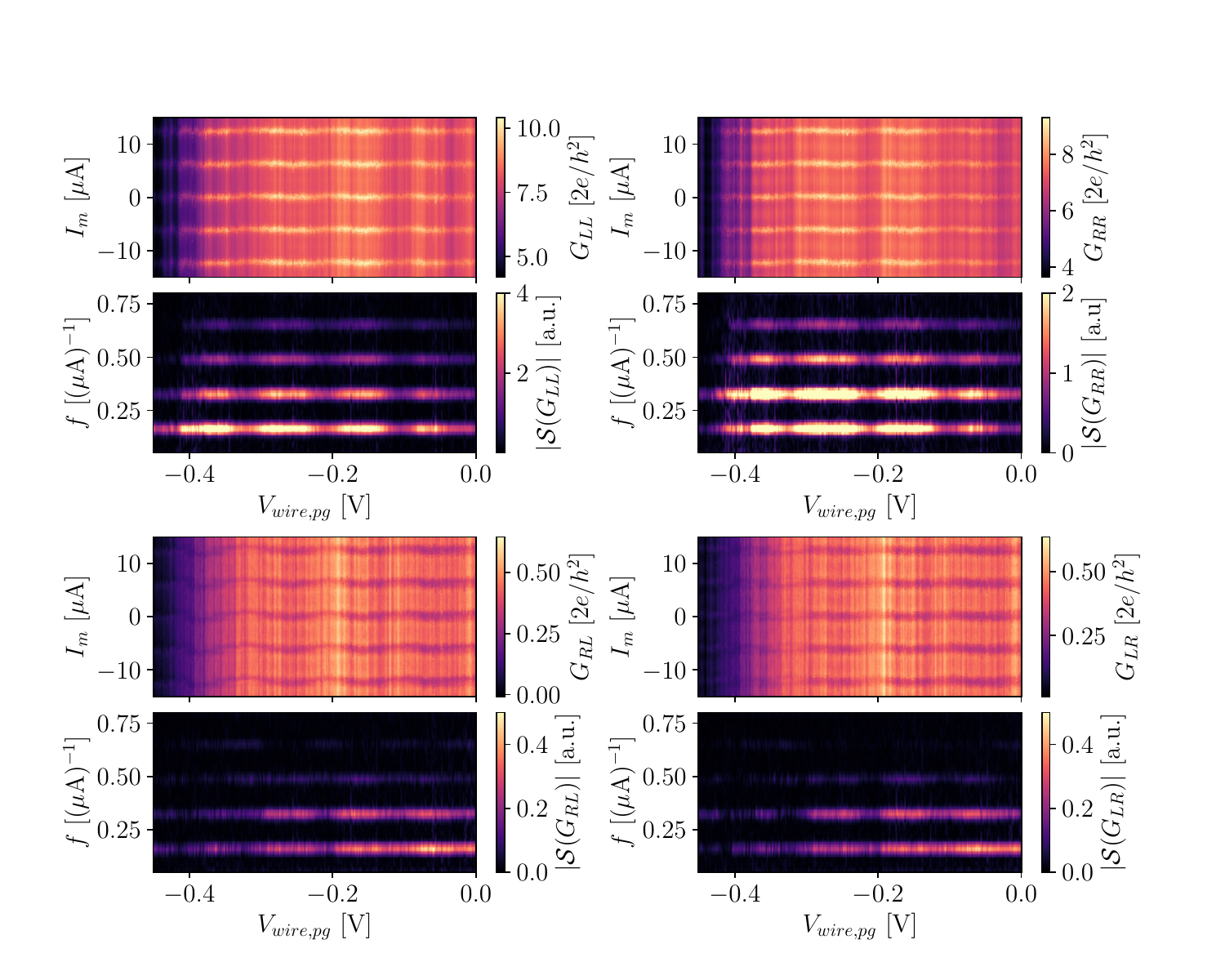}
\caption{Ten-wire device: Local conductance, $G_{LL/RR}$, and nonlocal conductance, $G_{LR/RL}$, as a function of the current through the meander, $I_{m}$, and voltage, $V_{wire,pg}$, on the gate between the wires. The lower part of each panel show the absolute value of the spectral density, $|S|$, as a function of frequency, $f$, associated with $I_{m}$.\label{figsup:4}}
\end{figure}

We have also varied the gates, $V_{wire,pg}$, covering the junctions between the wires of the 2, 3, 4 and 10 wire device, see \cref{figsup:1,figsup:2,figsup:3,figsup:4}. The behaviour observed in the main text remains: The periodicity is fixed, and principal and secondary maxima remains. We do observe some resonances, which likely originates from a cross-talk between the $V_{wire,pg}$ gate located on top of the junctions, and the potential landscape of the constrictions into the diffraction grating. The local and nonlocal signal is plotted with each fast Fourier transform (FFT). Around pinch off we have had to linearly interpolate data points in order to perform the FFT. To prepare the data for FFT, we have applied a gaussian window function and zero padding. Additionally the minimum conduction at each $V_{wire,pg}$ values has been subtracted from the signal in the FFT. The FFT tool is that provided by the SciPy library \cite{SciencePy}.

\begin{figure} 
\centering
\includegraphics[width=0.9\textwidth]{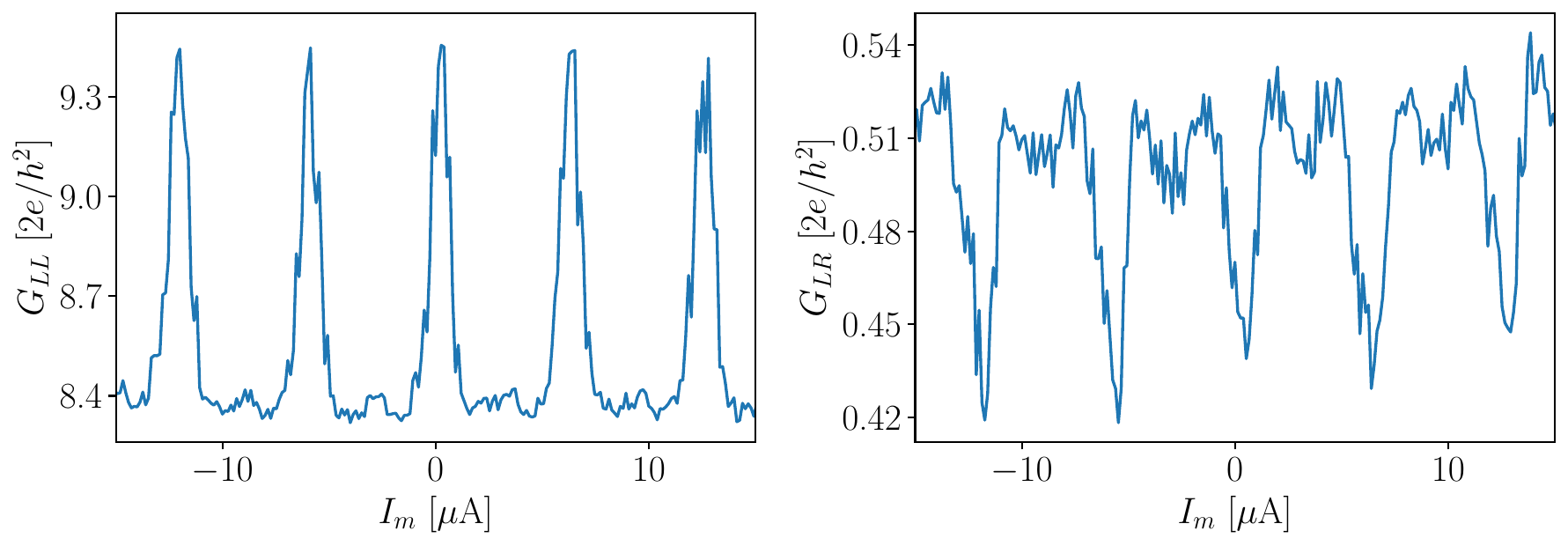}
\caption{Ten-wire device: Local conductance, $G_{LL}$, and nonlocal conductance, $G_{LR}$, as a function of the current through the meander, $I_{m}$, at $V_{wire,pg}=-0.2~$~V, from Fig.~\ref{figsup:4} Though we do not see the appearance of 9 secondary maxima, the primary peak is considerably narrower than fewer wire devices. \label{figsup:5}}
\end{figure}

\clearpage

\end{document}